\documentclass[conference]{IEEEtran}
\usepackage{graphicx}
\usepackage{epstopdf} 
\usepackage{verbatim} 
\usepackage{multirow} 
\usepackage{multicol}
\usepackage{amsfonts}
\usepackage{bbm}
\usepackage{subfigure}
\usepackage{colortbl} 
\usepackage{indentfirst}
\usepackage{dcolumn,booktabs}
\usepackage{algorithm}
\usepackage[noend]{algpseudocode}
\usepackage{algorithmicx,algorithm}

\usepackage{algpseudocode}
\usepackage{graphics}
\usepackage{epsfig,color}
\usepackage{amsmath}
\usepackage{amssymb}
\ifCLASSINFOpdf

\else
\fi
\newtheorem{Theorem}{Theorem}
\newtheorem{Corollary}{Corollary}

\newtheorem{Lemma}{Lemma}

\begin{document}

\title{Edge Caching with Real-Time Guarantees}

\author{
\IEEEauthorblockN{Le Yang\IEEEauthorrefmark{1}, Yi Zhong\IEEEauthorrefmark{2}, Fu-Chun Zheng\IEEEauthorrefmark{1}, Shi Jin\IEEEauthorrefmark{1}}
\IEEEauthorblockA{\IEEEauthorrefmark{1}National Mobile Communications Research Laboratory, Southeast University\\
\IEEEauthorrefmark{2}Huazhong University of Science and Technology, China
\\E-mails: \{yangle, jinshi\}@seu.edu.cn, yzhong@hust.edu.cn, fzheng@ieee.org}
}
\maketitle

\begin{abstract}
In recent years, optimization of the success transmission probability in wireless cache-enabled networks has been studied extensively. However, few works have concerned about the real-time performance of the cache-enabled networks. In this paper, we investigate the performance of the cache-enabled with real-time guarantees by adopting the age of information (AoI) as the metric to characterize the timeliness of the delivered information. We establish a spatial-temporal model by utilizing the stochastic geometry and queuing theory which captures both the temporal traffic dynamics and the interferers' geographic distribution. Under the random caching framework, we achieve the closed-form expression of AoI by adopting the maximum average received power criterion for the user association. Finally, we formulate a convex optimization problem for the minimization of the Peak AoI (PAoI) and obtain the optimal caching probabilities by utilizing the Karush-Kuhn-Tucker (KKT) conditions. Numerical results demonstrate that the random caching strategy is a better choice rather than the most popular content and uniform caching strategy to improve the real-time performance for the cached file as well as maintaining the file diversity at a relatively high level.
\end{abstract}

\begin{IEEEkeywords}
Age of information, Caching policy, Cache-enabled network, Stochastic geometry.
\end{IEEEkeywords}

%
\IEEEpeerreviewmaketitle

\section{Introduction}
Motivated by the significant development of the telecommunications technology and ubiquitous connectivity, numerous applications, such as weather forecasting, environment monitoring and instant messaging, are constantly emerging and has been widely used. In these applications, real-time status updates are becoming more popular and the timeliness of these messages is important. In this regard, age of information (AoI) is proposed to measure the freshness of the information and the age is defined as the difference between the current time and the generation time of the observed packet. Unlike the traditional metrics such as the delay or the throughput, an optimal rate can be determined by minimizing the AoI in order to ensure that the update status is as timely as possible at the receiver.

A considerable number of researches have been conducted on the AoI analysis under different queuing systems and service disciplines [1-5]. Furthermore, the AoI is introduced as a performance metric of the wireless network and the real-time characteristics of information delivery for wireless network is considered by utilizing AoI as a metric. The optimal scheduling policy is provided in order to minimize the weighted sum AoI of the single-source-multi-server queuing system in \cite{weight-sum}. In addition, virtue-queue and age based policies are proposed when the perfect channel side information is available \cite{perfect-CSI} or not \cite{not-perfect-CSI}.

However, the departure process of the packets are modeled as the poisson process and the spatial characteristics of the distance-dependent interferences are not considered. Since the stochastic geometry has been proved to be a powerful tool for evaluating the performance of wireless network where the node locations can be modeled as poisson point process (PPP). In \cite{AoI-network}, a decentralized scheduling policy is proposed to achieve the minimal AoI by taking the effect of interferences into consideration. The authors in \cite{AoI-network-zhong} discuss the AoI of the Geo/G/1 queuing system with deadline and provide the upper and lower bounds of the AoI.

Recently, proactively caching the popular files at the BSs or the edge servers has become a promising method to alleviate the burden of backhaul at on-peak traffic time and effectively reduce the access delay. Henceforth, we introduce the file caching into the AoI analysis. We take the caching of the dynamic map data in the cache-enabled roadside unit (RSU) as an example. The dynamic map data of a specific geographical area in the server can be regarded as a file. In order to efficiently reduce the AoI, the files are delivered to the BSs or the edge servers within the area. In addition, the popularity of each file is identical to all the BSs or the edge servers within the area. In such a scenario, the real-time acquisition of the map data is concerned by the drivers. In this regard, the files can be cached in the edge servers or BSs in advance and an appropriate content placement strategy to guarantee the real-time performance of the cache-enabled network is important and challenging.

In this paper, we establish a spatial-temporal model by utilizing the stochastic geometry and queuing theory. Under the random caching framework, we achieve the closed-form expression of AoI by adopting the maximum average received power criterion for the user association. Finally, we formulate a convex function for the minimization of the Peak AoI (PAoI) and obtain the optimal caching probabilities by utilizing the Karush-Kuhn-Tucker (KKT) conditions.

\section{System model}
We consider a downlink single-tier cache-enabled wireless network. The base stations (BSs) and the users are assumed to follow independent homogeneous Poisson Point Process (PPP) with density $\lambda$ and $\lambda_u$, respectively. Let $P$ denote the transmit power of each BS. The available bandwidth is denoted by $W$ (in Hz). The channel is subjected to Rayleigh fading and the power fading coefficient $h\sim\exp(1)$. We consider the path loss undergoes power law attenuation, i.e., the signal from a BS located at the distance $x$ undergoes the path loss $l(x)=x^{-\alpha}$, where $\alpha$ denotes the path loss exponent. Without loss of generality, according to Slivnyak's theorem \cite{Bai}, we can study the performance of a typical user $u_0$ equipped with a single antenna located at the origin $o\in\mathbb{R}$. Moreover, we consider the discrete system where the time is divided into discrete time slots.

Apart form the spatial model which is stated aforementioned, we introduce the temporal arrival model for the BSs. The packets arrival process for $u_0$ is modeled as a Bernoulli process with the arrival rate $\zeta$, i.e., the time between two adjacent arrivals is subject to the geometric distribution with parameter $\zeta$. The transmission process is considered to be successful if the signal-to-interference-plus-noise ration (SINR) exceed a certain threshold $\theta$. The packet is removed out of the buffer when a ACK message is received from the user. If the transmitted packet is not decoded correctly, a NACK feedback is transmitted by the user and the packet will be retransmitted.

The random nature of the traffic affects the performance of the cache-enabled network greatly. In order to reduce the influences of the interferences as well as improving the overall performance of the cache-enabled network, we apply the mechanism where the BS is muted randomly in each time slot. No data is transmitted in the muted time slots and the active probability for each BS is $\beta$.

Let $\mathcal{F}\triangleq \{1,2,\cdots,F\}$ denote a set of $F$ files in the network. For analytical tractability, all files are assumed to have the same size, and the file popularity distribution is identical among all users. Let $p_f$ denote the probability that File $f$ is requested by a user, i.e., the popularity of File $f$ is $p_f$, where $\sum_{f=1}^{F}p_f=1$. In addition, we assume that $p_1\ge p_2 \ge \cdots \ge p_F$. Hence, the file popularity distribution can be expressed as $\mathbf{p}\triangleq \{p_1,p_2,\cdots,p_F\}$, which is assumed to be known a priori. In addition, The popularity of different files evolves at a relatively slow timescale and can be estimated by utilizing machine learning techniques. Each BS is equipped with a cache of size $C \le F$ to store $C$ different files from $\mathcal{F}$.

\subsection{Random Caching}

To provide the spatial diversity and improve the system performance, the random caching strategy \cite{cui} is adopted at different tiers. Let $\mathbf{q}\triangleq\{q_1, q_2,\cdots,q_F\}$ denote the caching probability matrix where $q_{f}$ denotes the caching probability of File $f$. Then, we have
\begin{eqnarray}\label{probability constraint}
&& 0\le q_{f} \le 1,\\
&& \sum\limits_{f=1}^{F}q_{f}=C,\label{capacity constraint}
\end{eqnarray}
which indicates that the sum of all the caching probabilities of the files cached in the same tier is limited by the BS storage capacity of the corresponding tier.

Let $\Phi_{t,f}$ denote the set of BSs caching File $f$ at time slot $t$. Consider $u_0$ requesting File $f$. If File $f$ is not stored in any tier, $u_0$ will download the corresponding file from the core network through the backhaul links. Otherwise, $u_0$ is associated with the BS which not only stores File $f$ but also provides the maximum average received signal strength among all BSs in the HetNet, referred to as the serving BS.

We consider the interference-limited regime and thus the thermal noise is neglected. The SINR can be written as follows
\begin{equation}
\text{SINR}=\frac{Ph_0r^{-\alpha}}{I_{t,f}+I_{t,-f}},
\end{equation}
where the interferences from the BSs caching File $f$ at time slot $t$ $I_{t,f}=\sum_{i\in\Phi_{t,f}\backslash x_0}Ph_ix_i^{-\alpha}$, the interferences from the BSs not caching File $f$ at time slot $t$ $I_{t,-f}=\sum_{i\in\Phi_{t,-f}}Ph_ix_i^{-\alpha}$.
Therefore, the successful transmission probability (STP) is given by
\begin{equation}
\text{Pr}(\theta)=\beta\mathbbm{P}\left[\text{SINR}>\theta\right]
\end{equation}

\subsection{Performance Metric}
Fig. 1 shows the variation of the AoI as a function of the time. We assume that the observation starts at time 0 when the age is set to 1. Afterwards, the desired file is retrieved from the storage space of the BSs and waited to be transmitted at time $t_k,k\in\{1,2,\cdots,K\}$, where $K$ denotes the index of the recently retrieved file. The corresponding file is received by $u_0$ successfully at time $t_k^{'}$. Specifically, the evolution process of the AoI can be divided into two conditions. the AoI grows linearly if the transmission of the desired file fails. Otherwise, the AoI is reset to the time elapsed since the instant when the desired file arrivals at the queue of the serving BS, i.e., $t_k-t_k^{'}$. To make the description more precise, the AoI evolution process can be written as follows
\begin{equation}
A(t_{k+1})=
\begin{cases}
A(t_k)+1\\
t_k-t_k^{'}+1
\end{cases}
\end{equation}

\begin{figure}
  \centering
  \includegraphics[width=3.5in]{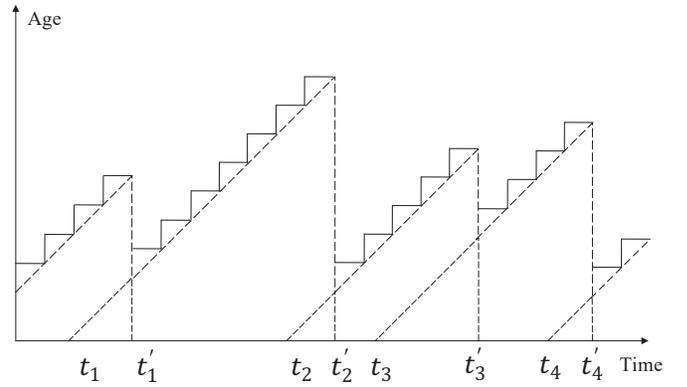}
  \caption{The evolution process of AoI.}\label{fig.sim.se}
\end{figure}

Depending on the demand of the applications, it may be desirable to ensure that the AoI remains below a threshold. Hence, we utilize the peak AoI (PAoI) as the performance metric, which provides the worst case age. Consider the PAoI of the $k$th successfully delivered file which is denoted by $A_k$. $A_k$ can be expressed as follows
\begin{equation}
\mathbbm{E}[A_k|\Phi]=T_k+Y_k,
\end{equation}
where $Y_k$ denotes the inter-arrival time between the $k$th and the $k-1$th packets. Since the arrival traffic is a Bernoulli process, we have $\mathbbm{E}(T_k)=1/\zeta$. In addition, $Y_k$ denotes the sojourn time for the $k$th packet. The expectation of $Y_k$ for the Geo/G/1 queue can be written as
\begin{equation}
\mathbbm{E}[T_k]=
\begin{cases}
\frac{1-\zeta}{\mu-\zeta},\quad \text{if}\ \mu>\zeta\\
\infty,\quad \text{if}\ \mu>\zeta
\end{cases}
\end{equation}

\section{AoI Analysis}
In this section, we first achieve the closed-form expression of the PAoI under the general case, then provide the simplified expression of the PAoI under the special case where the success transmission probability is far larger than the arrival rate.
\begin{Theorem}
In the cache-enabled networks where each BS is active independently with probability $\beta$, the PAoI is given by
\begin{equation}\label{theorem}
\begin{split}
&\mathbbm{E}[A_k]=\frac{1}{\zeta}+\sum_{f=1}^{F}p_f(1-\zeta)q_f
\left(q_f-\sum_{n=0}^{\infty}\zeta^{n}\beta^{-n-1}\delta\right.\\
&\left.\sum_{m=1}^{n+1}\binom{n+1}{m}\left(\frac{(\beta\theta)^{m}q_f}{m-\delta}W(\theta)
+(1-q_f)\beta^m\theta^{\delta}V(\theta)\right)\right)^{-1}
\end{split}
\end{equation}
where $W(\theta)={}_{2}F_{1}(m,m-\delta;m-\delta+1;-(1-\beta)\theta)$, $V(\theta)=\text{B}(\delta,m-\delta)$. Note that ${}_{2}F_{1}(\cdot)$ denotes Gauss hypergeometric function and $\text{B}(\cdot)$ denotes the Beta function.
\end{Theorem}
\emph{Proof:}
Assuming that $u_0$ requests File $f$, the success transmission probability conditioned on the realization $\Phi$ can be derived as follows
\begin{equation}
\begin{split}
\text{Pr}(\theta)&\overset{(a)}{=}\beta\text{Pr}\left(\text{SIR}_{k,t}>\theta|\Phi\right)\\
&=\beta\text{Pr}\left(Ph_0r^{-\alpha}>\theta(I_{t,f}+I_{t,-f})|\Phi\right)\\
&\overset{(b)}{=}\beta\mathbbm{E}_{I}\left[\exp\left(-\frac{\theta r^{\alpha}}{P}I_{t,f}\right)
\exp\left(-\frac{\theta r^{\alpha}}{P}I_{t,-f}\right)\right]\\
&\overset{(c)}{=}\beta\mathbbm{E}_{I}\left[\mathcal{L}_{I_{t,f}}\left(\frac{\theta r^{\alpha}}{P}\right)
\mathcal{L}_{I_{t,-f}}\left(\frac{\theta r^{\alpha}}{P}\right)\right]
\end{split}
\end{equation}
where (a) follows from the definition of $\text{Pr}(\theta)$, (b) follows from that $h_i$ is subjected to the exponential distribution with unit mean. In the last step of the above derivation, $\mathcal{L}_{I_{t,f}}\left(\frac{\theta r^{\alpha}}{P}\right)$ and $\mathcal{L}_{I_{t,-f}}\left(\frac{\theta r^{\alpha}}{P}\right)$ denote the Laplace transforms of the interferences of the BSs with/without caching File $f$, respectively. Assuming $s=\frac{\theta r^{\alpha}}{P}$, the Laplace transform is achieved in the following. Firstly, the Laplace transform of the interferences from the BSs caching File $f$ can be evaluated as
\begin{equation}\label{laplace-f}
\begin{split}
\mathcal{L}_{I_{t,f}}(s|\Phi)&=\mathbbm{E}_{I_{t,f}}\left[\exp\left(-sI_{t,f}\right)\right]\\
&=\mathbbm{E}_{h_{i}}\left[\exp\left(-s\sum_{i\in\Phi_{t,f}\backslash x_0}Ph_ix_i^{-\alpha}\right)\right]\\
&=\prod_{i\in\Phi_{t,f}\backslash x_0}\mathbbm{E}_{h_{i}}\left[\beta\exp\left(-sPh_ix^{-\alpha}\right)+1-\beta\right]\\
&=\prod_{i\in\Phi_{t,f}\backslash x_0}\left(\frac{\beta}{1+sPx_i^{-\alpha}}+1-\beta\right)
\end{split}
\end{equation}

Similarly, the Laplace transform of the interferences from the BSs not caching File $f$ can be evaluated as
\begin{equation}\label{laplace-no-f}
\begin{split}
\mathcal{L}_{I_{t,f}}(s|\Phi)
=\prod_{i\in\Phi_{t,-f}}\left(\frac{\beta}{1+sPx_i^{-\alpha}}+1-\beta\right)
\end{split}
\end{equation}

Therefore, the conditional PAoI can be derived as follows
\begin{equation}\label{AoI}
\begin{split}
&\mathbbm{E}[A_k|f]\\
&=\mathbbm{E}_{\Phi}\left[\frac{1}{\zeta}+\frac{1-\zeta}{\text{Pr}_{\Phi,f}-\zeta}\right]\\
&\overset{(a)}{=}\mathbbm{E}_{\Phi}\left[\frac{1}{\zeta}+\frac{1-\zeta}{\text{Pr}_{\Phi,f}}\sum_{n=0}^{\infty}\left(\frac{\zeta}{\text{Pr}_{\Phi,f}}\right)^n\right]\\
&\overset{(b)}{=}\mathbbm{E}_{\Phi}\left[\frac{1}{\zeta}+(1-\zeta)\sum_{n=0}^{\infty}\frac{\zeta^{n}}{\beta^{n+1}}
\left(\frac{1}{\mathcal{L}_{I_{t,f}}(s)\mathcal{L}_{I_{t,-f}}(s)}\right)^{n+1}\right]\\
\end{split}
\end{equation}
where (a) utilizes the Taylor series, i.e., $\frac{1}{1-x}=\sum_{n=0}^{\infty}x^n$. By substituting (\ref{laplace-f}) and (\ref{laplace-no-f}) into (\ref{AoI}), and utilizing the probability generating functional (PGFL) of the PPP, the AoI can be obtained in (\ref{AoI-final}), as shown on the top of the next page.
\begin{figure*}
\begin{equation}\label{AoI-final}
\begin{split}
\mathbbm{E}[A_k|f]=&\mathbbm{E}_{\Phi}\left[\frac{1}{\zeta}+(1-\zeta)\sum_{n=0}^{\infty}\frac{\zeta^{n}}{\beta^{n+1}}
\exp\left(\pi\delta\lambda q_f\sum_{m=1}^{n+1}\binom{n+1}{m}\frac{(\beta\theta)^mr^2}{m-\delta}{}_{2}F_{1}(m,m-\delta;m-\delta+1;-\theta)\right.\right.\\
&+\left.\left.\pi\delta\lambda(1-q_f)\sum_{m=1}^{n+1}\binom{n+1}{m}\beta^m\theta^{\delta}r^2B(\delta,\delta-m)\right)
\right]\\
\end{split}
\end{equation}
\end{figure*}

According to \cite{distance-pdf}, the probability density function (PDF) of the distance between $u_0$ and the serving BS conditioned on that File $f$ is requested is given by
\begin{equation}\label{pdf}
f_R(r)=2\pi\lambda q_{f}r\exp(-\pi\lambda q_fr^2)
\end{equation}

Therefore, averaging over $r$ yields that
\begin{equation}
\begin{split}
&\mathbbm{E}[A_k|f]=\frac{1}{\zeta}+\int_{0}^{\infty}2\pi\lambda q_f(1-\zeta)r\sum_{n=0}^{\infty}\frac{\zeta^{n}}{\beta^{n+1}}\\
&\exp\left(-\pi\lambda q_fr^2+\pi\delta\lambda q_fr^2\sum_{m=1}^{n+1}\binom{n+1}{m}\frac{(\beta\theta)^mr^2}{m-\delta}W(\theta)\right.\\
&\left.+\pi\delta\lambda(1-q_f)r^2\sum_{m=1}^{n+1}\binom{n+1}{m}\beta^m\theta^{\delta}r^2V(\theta)\right)\text{d}r
\end{split}
\end{equation}

Since $\int_{0}^{\infty}2re^{-Ar^2}=1/A$, the results in (\ref{theorem}) can be obtained.

From Theorem 1, it is observed that PAoI is affected by two sets of parameters: the physical layer parameters and the content-related parameters. We can see that PAoI is a decreasing function of the path loss exponent $\alpha$. This can be explained as follows. When $\alpha$ increases, the interferences attenuates faster and the SIR increases accordingly. Note that PAoI is independent of the density $\lambda$, which indicates that the PAoI remains unchanged when adding more BSs into the cache-enabled network. A intuitive explanation can be provided. When the density increases, the average distance between $u_0$ and the serving BS becomes shorter and the received signal becomes stronger. Meanwhile, the interferences experienced by $u_0$ is larger. Therefore, the statistics of the SIR does not change.

From (\ref{theorem}), we can see the PAoI is the increasing function of the SINR threshold $\theta$. The explanation can be stated as follows. When $\theta$ increases, the STP decreases indicating that larger amounts of time slots are needed in order to transmit the files successfully. Therefore, the PAoI is increased accordingly. From (\ref{PAoI}), we obtain that $\mathbbm{E}(A_k)\rightarrow\infty$ when $\text{Pr}(\theta)<\zeta$. That is, when the physical layer parameters and the arrival rates are given, the PAoI approaches infinity when $\theta$ exceeds a certain value. we define the critical value of $\theta$ as
\begin{equation}
\theta_c\triangleq\sup\{\theta:\text{Pr}(\theta)<\zeta\},
\end{equation}
which indicates that $\theta$ needs to maintain below $\theta_c$ in order to achieve a finite PAoI.

Moreover, the PAoI increases with decreasing caching probability. The reason is that the STP increases with the increasing caching probability and causes a degradation in the PAoI. In this regard, given $\theta$, the caching probabilities for all the files needs to maintain above a certain threshold $q_{f}^{c}$, which is written as follows
\begin{equation}
q_f^c\triangleq\inf\{q_f:\text{Pr}(\mathbf{q})<\zeta\}.
\end{equation}

The lower bound of caching probability for File $f$ $q_f^c$ can be computed easily as follows
\begin{equation}
q_f^c=\frac{E(\theta)}{1-G(\theta)+E(\theta)}
\end{equation}
where
\begin{equation}
G(\theta)=\sum_{n=0}^{\infty}\zeta^n\beta^{m-n-1}\delta\sum_{m=1}^{n+1}\binom{n+1}{m}\theta^m(m-\delta)^{-1}W(\theta)
\end{equation}
\begin{equation}
E(\theta)=\sum_{n=0}^{\infty}\zeta^n\beta^{m-n-1}\delta\sum_{m=1}^{n+1}\binom{n+1}{m}\theta^{\delta}V(\theta)
\end{equation}

We can observe that the PAoI is an increasing function of the SINR threshold $\theta$. It coincides with our intuition that successful transmission probability decreases when $\theta$ increases, thereby leading to the increasing PAoI. In addition, it is observed that
The expression of PAoI in the general case includes the sum of infinite terms with the appliance of the Taylor series expansion. Therefore, it take considerable time to obtain the numerical results with the acceptable accuracy. In the following corollary, we consider the condition where the success transmission probability is far larger than the arrival rate, i.e., $\text{Pr}(\theta)>\zeta$. Under this assumption, the expression of PAoI can be simplified significantly by neglecting the higher order terms.

\begin{Corollary}
When $\text{Pr}(\theta)>>\zeta$, the PAoI of the cache-enabled network is given by
\begin{equation}\label{corollary}
\begin{split}
&\mathbbm{E}[A_k]=\frac{1}{\zeta}+\sum_{f=1}^{F}(p_f(1-\zeta)q_f)\left(q_f-q_f\theta\frac{2}{\alpha-2}W_1(\theta)\right.\\
&\left.-(1-q_f)\theta^{\delta}V_1(\theta)\right)^{-1}+((1-\zeta)\zeta q_f)\left(q_f-q_f\theta\left(\frac{2}{\alpha-2}\right.\right.\\
&\left.\left.W_1(\theta)+\frac{\theta}{\alpha-1}W_2(\theta)\right)
-(1-q_f)\theta^{\delta}(V_1(\theta)+V_2(\theta))\right)^{-1}
\end{split}
\end{equation}
where $W_1(\theta)={}_{2}F_{1}(1,1-\delta;2-\delta;-(1-\beta)\theta)$, $W_2(\theta)={}_{2}F_{1}(2,2-\delta;3-\delta;-(1-\beta)\theta)$, $V_1(\theta)=\text{B}(\delta,1-\delta)$ and $V_2(\theta)=\text{B}(\delta,2-\delta)$.

\emph{Proof:}
Conditioned on $\text{Pr}>>\zeta$, (\ref{AoI}) can be simplified as
\begin{equation}\label{corollary-proof}
\begin{split}
&\mathbbm{E}[A_k|f]=\mathbbm{E}_{\Phi}\left[\frac{1}{\zeta}+\frac{1-\zeta}{\text{Pr}_{\Phi,f}}\left(1+\frac{\zeta}{\text{Pr}_{\Phi,f}}\right)\right]\\
&=\mathbbm{E}_{\Phi}\left[\frac{1}{\zeta}+\frac{1-\zeta}{\beta\mathcal{L}_{I_{t,f}}(s)\mathcal{L}_{I_{t,-f}}(s)}
+\frac{(1-\zeta)\zeta}{\beta^2\mathcal{L}_{I_{t,f}}^2(s)\mathcal{L}_{I_{t,-f}}^2(s)}\right]
\end{split}
\end{equation}
Substituting (\ref{laplace-f}), (\ref{laplace-no-f}) and (\ref{pdf}) into (\ref{corollary-proof}), the results in (\ref{corollary}) can be obtained.
\end{Corollary}

\section{AoI Optimization}
In this section, we optimize the caching probabilities for all the files cached in the network to obtain the minimum PAoI.

\emph{Problem 1(Optimization of PAoI):}
\begin{equation}
\min\limits_{\mathbf{q}}\mathbbm{E}[A_k]
\end{equation}
\emph{s.t.}
\begin{equation}
q_f^c\leq q_f\leq1,
\end{equation}
\begin{equation}
\sum_{f=1}^{F}q_f=C.
\end{equation}

Since the objective function and the constraint set are both convex, Problem 1 is a convex optimization problem. We therefore utilize the KKT conditions to obtain the globally optimal caching probabilities.

\begin{Lemma}
In the cache-enabled network where each BS is active independently with probability $\beta$, the optimal caching probability of File $f$ is given by
\begin{equation}
\begin{split}
q_{f}^{*}=&\min\left\{\max\left\{\frac{1-\zeta}{\beta(1-G(\theta)+E(\theta))}\right.\right.\\
&\left.\left.\sqrt{\frac{p_fE(\theta)}{\eta^*}}
+\frac{(1-\zeta)E(\theta)}{\beta(1-G(\theta)+E(\theta))},q_f^c\right\},1\right\}
\end{split}
\end{equation}

Note that $\eta^*$ satisfies
\begin{equation}
\begin{split}
&\sum_{f=1}^{F}\left(\frac{1-\zeta}{\beta(1-G(\theta)+E(\theta))}\sqrt{\frac{p_fE(\theta)}{\eta^*}}
+\frac{(1-\zeta)E(\theta)}{\beta(1-G(\theta)+E(\theta))}\right)\\
&=C
\end{split}
\end{equation}

\end{Lemma}

From Lemma 1, we can see that $\eta^{*}$ is jointly affected by the physical layer parameters and the content-related parameters, i.e., the file popularity $\mathbf{p}$ and the caching size $C$. Specifically, the physical layer parameters affect the caching probabilities of all files in the common manner while the file popularity $\mathbf{p}$ affects the file of the corresponding file.

\section{Simulation Results}
In this section,  We first illustrate the the impact of different physical layer parameters on the PAoI, then compare the optimal caching strategy with two benchmark strategies.

Two caching strategies are provided as the benchmark:

1) Most Popular Content (MPC): Caching the most popular files in each BS. No file diversity is achieved.

2) Uniform Caching (UC): Caching each file with equal probability. The maximal file diversity can be achieved.

We assume the popularity of the files satisfies the Zipf distribution, i.e., $p_{f}=\frac{f^{-\psi}}{\sum_{f=1}^{F}f^{-\psi}}$, where $\psi$ denotes the skew parameter. Unless otherwise stated, the parameter settings are listed as follows. The transmit power $P=23$dBm, the density is $\lambda=3/(250^2\pi)$, path loss exponent $\alpha=4.5$, the skewness of the file popularity $\psi=0.8$, the total number of files $N=30$.

In Fig. 2, it is observed that PAoI is an increasing function of the SINR threshold, which coincides with our theoretical analysis. Moreover, we can observe that although the PAoI with the lowest the arrival rate $\zeta$ is large at the start, it is the last one to approaches infinity, i.e., the critical value of the SINR threshold decreases with the increasing arrival rate $\zeta$. Intuitively, when the SINR threshold $\theta$ is small, the dominant factor which affects the PAoI is the arrival rate $\zeta$ rather than the STP. Therefore, the PAoI with larger arrival rate becomes prominently larger. However, the PAoI with lower arrival rate $\zeta$ could keep the finite value even in the relative large SINR threshold regime. This is because that when the arrival rate $\zeta$ is relatively low, the SINR threshold $\theta$ needs to be large enough to make the STP approach the arrival rate $\zeta$. In contrast, the SINR threshold $\theta$ can maintain at a relative small value to satisfy the condition where the STP approaches the arrival rate when the arrival rate $\zeta$ is large.

\begin{figure}
  \centering
  \includegraphics[width=3.5in]{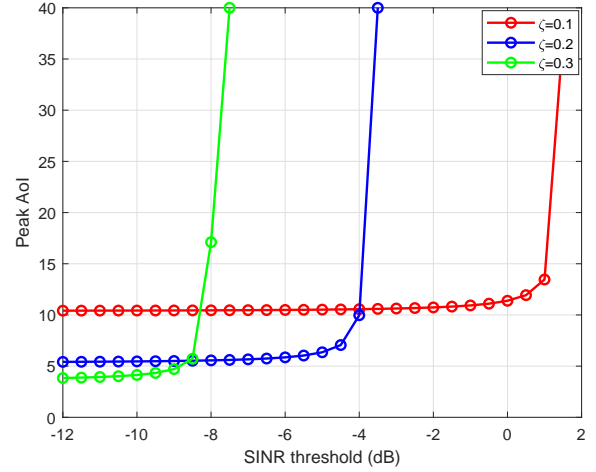}
  \caption{PAoI versus SINR threshold under different arrival rates.}\label{fig.sim.se}
\end{figure}

Fig. 3 shows the PAoI as the function of the cache size under different caching strategies. The PAoI decreases significantly when the cache size is increased under the UC strategy. The random caching strategy can achieve the lower PAoI compared with the UC strategy. The is because that the increasing cache size enlarges the caching probability for each file and improves the corresponding STP. In addition, we find that the gap between the UC strategy and the random caching strategy is larger with small cache size. This provide a useful guideline that optimizing the caching probability for each file can better reduce the PAoI when the cache size is relatively small.

Note that the PAoI increases with the increasing cache size under the MPC strategy. The reason is that only the cached file is considered while evaluating the PAoI and the PAoI increases when more files is cached in the BSs. Actually, we need to retrieve the files not cached in the BSs from the core network through the backhaul and the delay is considerable. According to whether taking the backhaul cost caused by retrieving the file not cached in the BSs into consideration or not, different caching strategies should be adopted. If we take the backhaul cost into consideration, we have better adopt the UC strategy since it achieves the maximal file diversity. Otherwise, we should adopt the MPC strategy since the files are cached in each BSs and the best real-time performance can be achieved with the cached files. In order to achieve the good real-time performance as well as maintaining the file diversity, the random caching strategy is the appropriate choice.

\begin{figure}
  \centering
  \includegraphics[width=3.5in]{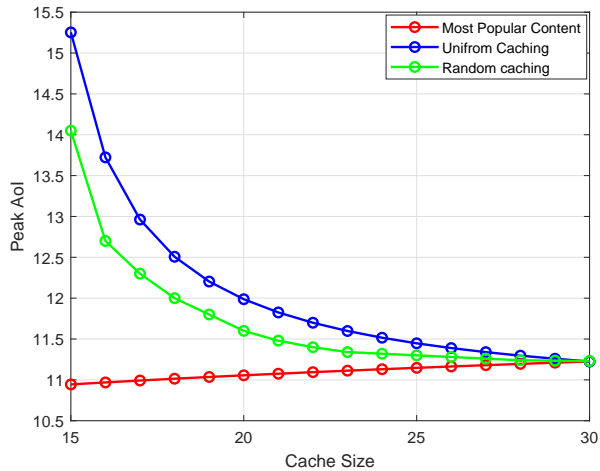}
  \caption{PAoI versus cache size under different caching strategies.}\label{fig.sim.se}
\end{figure}

\section{Conclusion}
In this paper, we investigated the real-time performance of the cache-enabled network by adopting the AoI as the metric to characterize the timeliness of the delivered information. In addition, we adopted the random mute mechanism to effectively reduce the effect of the interferences. Under the random caching framework, we considered that the user was associated with the BS under the maximum received power criterion and achieved the closed-form expression of the PAoI. We formulated the convex optimization problem and obtained optimal caching probabilities by utilizing the Karush-Kuhn-Tucker (KKT) conditions. Moreover, the numerical results provided some useful insight into the impact of the physical layer parameters and the arrival rate on the AoI. In addition, Numerical results demonstrate that the random caching strategy is a better choice rather than the most popular content and uniform caching strategy to improve the real-time performance for the cached file as well as maintaining the file diversity at a relatively high level.

\end{document}